\documentclass[english]{article}
\usepackage[T1]{fontenc}
\usepackage[latin1]{inputenc}
\usepackage{graphicx}
\usepackage{authblk}
\makeatletter

\usepackage{babel}

\title{Gravitational Lensing by Wormholes}

\author[1]{Juan Manuel Tejeiro S.}

\author[2]{Eduard Larranaga}

\affil[1,2]{\emph{Observatorio Astronomico Nacional. Universidad Nacional de
Colombia}}

\begin{document}
\maketitle

\begin{abstract}
Natural wormholes and its astrophysical signatures have been sugested
in various oportunities. By applying the strong field limit of gravitational
lensing theory, we calculate the deflection angle and magnification
curves produced by Morris-Thorne wormholes in asimptotically flat
space-times. The results show that wormholes act like convergent lenses.
Therefore, we show that it is hard to distinguish them from black
holes using the deflection's angle of the gravitational lens effect,
in contrast with the results reported by Cramer et.al. and Safanova
et.al. However, we also show that it is possible, in principle, distinguish
them by the magnification curves, in particular, by observing the
position of the peak of the Einstein's ring.
\end{abstract}

\section{Introduction}

Wormholes are space-time regions with two mouths connected by a throat.
Since the mouths are not hidden by event horizons and there is no
singularity inside, wormholes permit the passage of massive particles
or photons from one side to the other. Morris and Thorne\cite{Morris1,Morris2}
shown that wormholes require the violation of the averaged null energy
condition in order to satisfy the Einstein field equations in the
throat. Thus, matter in this region must exert gravitational repulsion
to make a stable configuration. The study of these solutions lead
to an interesting question: May some observable electromagnetic signatures
give the possibility of wormhole identification?\\

One of the most important applications of General Relativity is Gravitational
Lensing. Since matter inside the wormhole antigravitate, Cramer et.
al. \cite{cramer} and Safanova et. al. \cite{safanova} have studied
the lensing effects of negative masses on light rays from point sources
in the background, and argumented that this will be the effect produced
by natural stellar-size wormholes. The study of lensing effects show
that positive masses acts like convergent lenses, while the study
of Safanova et. al. for negative masses show that these act basically
like divergent lenses. This means that the deflection angle is negative
(divergent). Therefore, light from a point source will be concentrated
on the border of an \emph{umbra region}, producing two intensity enhancements:
one before and one after the occultation event.\\

Recently, the scientific community started to look gravitational lensing
from a strong field perspective. This limit refers to the effect produced
by a very massive object and/or when the light rays that are affected
pass very close to the lens. Bozza \cite{bozza2} provided an analytic
method to calculate the deflection angle for any spherically symmetric
spacetime in the strong field limit, and one important result of this
is that a light ray with a very small impact parameter will wind one
or several times around the lens before emerging, producing an infinite
set of \emph{relativistic images}.\\

In this paper, we will use the method of Bozza to calculate a more
realistic lensing effect of Morris-Thorne wormholes, and it will permit
us to show that wormhole spacetimes present a \emph{positive deflection
angle}, i.e. wormholes does not act like divergent lenses in the general
case. This means that exotic matter needed to construct the solutions
does not have a direct observable signature for an outside observer.

In section II we give the metric of the studied wormhole solutions,
section III gives the essential knowledge about strong gravitational
lensing as described by Bozza and in section IV we calculate the deflection
angle for the presented wormholes. In section V we compare the obtained
angles with the Schwarzschilds deflection angle and finally, in section
VI we put some conclusions.

\section{Morris-Thorne Wormholes}

In the pionnering work of Morris and Thorne they describe a general
class of solution of Einstein equations representing wormholes. The
conditions that they impose in order to obtain the general form of
the metric tensor include: \smallskip{}

1. The metric must be spherically symmetric and static (independient
of time)

2. The solution must have a throat joining two asymptotically flat
regions of spacetime

3. The metric must satisfy Einstein equations in every point of spacetime

4. The geometry will not have event horizons nor singularities \smallskip{}

Under these, the general form for the metric tensor is

\begin{equation}
ds^{2}=-e^{2\Phi\left(r\right)}c^{2}dt^{2}+\frac{1}{1-\frac{b\left(r\right)}{r}}dr^{2}+r^{2}\left(d\theta^{2}+\textrm{sin}^{2}\theta d\phi^{2}\right),\end{equation}

where $\Phi\left(r\right)$ is the redshift function and $b\left(r\right)$
is the form function. When imposing condition 3 above, we obtain the
relation between the functions $\Phi\left(r\right)$ and $b\left(r\right)$
and the stress-energy tensor that produces the wormhole spacetime
geometry. Outside the wormhole, the spacetime is considered asimptotically
flat and have a line element given by Schwarzschild metric,

\begin{equation}
ds^{2}=-\left(1-\frac{2GM}{r^{2}}\right)dt^{2}+\left(1-\frac{2GM}{r^{2}}\right)^{-1}dr^{2}+r^{2}\left(d\theta^{2}+\textrm{sin}^{2}\theta d\phi^{2}\right).\end{equation}

The junction conditions that follow from the theory of general relativity
are the continuity of the metric components and the extrinsic curvature
across the surface of juncture. \smallskip{}

To obtain specific solutions, in wormhole studies is usual to fix
a spacetime geometry (i.e. to fix the functions $\Phi$ and $b$ in
order to obtain a wormhole geometry) and then use the Einstein equations
to derive the matter distribution nedded to obtain the respective
metric. Folowing \cite{Morris1,Morris2} is easy to see that many
wormhole geometries need \emph{exotic matter} (i.e. matter that does
not obey the null energy condition) to exist, and it is localizated
at the throat of the wormhole. This need is what led Cramer et. al.
\cite{cramer} and Safanova et. al. \cite{safanova} to model a wormhole
as a point of negative mass to study gravitational lensing.

\subsection{Specific Solutions}

In this paper we will consider only two specific wormhole solutions
obtained by Lemos et. al. \cite{Lemos}. In both cases we will define
$r_{m}$ as the radius of the throat and $a$ as the radius of the
mouth of the wormhole.

\subsubsection{Solution A}

The first solution has an interior metric ($r_{m}\leq r\leq a$) given
by

\begin{equation}
ds^{2}=-\left(1-\sqrt{\frac{r_{m}}{a}}\right)dt^{2}+\frac{dr^{2}}{\left(1-\sqrt{\frac{r_{m}}{r}}\right)}+r^{2}\left(d\theta^{2}+\textrm{sin}^{2}\theta d\phi^{2}\right),\end{equation}

and an exterior metric ($a<r<\infty$) given by

\begin{equation}
ds^{2}=-\left(1-\frac{\sqrt{r_{m}a}}{r}\right)dt^{2}+\frac{dr^{2}}{\left(1-\frac{\sqrt{r_{m}a}}{r}\right)}+r^{2}\left(d\theta^{2}+\textrm{sin}^{2}\theta d\phi^{2}\right).\end{equation}

Since the exterior solution must represent Schwarzschild metric, we
can associate a mass to the wormhole, \begin{equation}
M=\frac{\sqrt{r_{m}a}}{2G},\end{equation}

and as can be seen , the complete metric is smoothly joined at mouth
radius $a$.

\subsubsection{Solution B}

The second solution has an interior metric ($r_{m}\leq r\leq a$)
given by

\begin{equation}
ds^{2}=-\left(1-\left(\frac{r_{m}}{a}\right)^{2}\right)dt^{2}+\frac{dr^{2}}{\left(1-\left(\frac{r_{m}}{r}\right)^{2}\right)}+r^{2}\left(d\theta^{2}+\textrm{sin}^{2}\theta d\phi^{2}\right),\end{equation}

and an exterior metric ($a<r<\infty$) given by

\begin{equation}
ds^{2}=-\left(1-\frac{r_{m}^{2}}{ar}\right)dt^{2}+\frac{dr^{2}}{\left(1-\frac{r_{m}^{2}}{ar}\right)}+r^{2}\left(d\theta^{2}+\textrm{sin}^{2}\theta d\phi^{2}\right).\end{equation}

Again, the metric is smoothly joined at mouth radius $a$, and since
the exterior solution must represent Schwarzschild metric, we associate
a mass to the wormhole, \begin{equation}
M=\frac{r_{m}^{2}}{2aG}.\end{equation}

Both metrics represent wormholes connecting two asimptotically flat
spacetimes, and to both of them we associate a positive external mass.
This fact inspire us to think that a light ray that passes near the
wormhole but not into it, will have a deflection similar to that produced
by a Schwarzschild geometry, while a light ray that goes into the
wormhole will have a little different behavior.

\section{Gravitational Lensing}

Because of the arguments given in the preceding section, the gravitational
lensing effects produced by a wormhole will be similar to those produced
by a Schwarzschild metric when the deflected light ray does not enter
into the wormhole, but in order to obtain a mathematical expresion
for the deflection angle of a light ray passing very close to the
wormhole we will use the analytical method for strong field limit
gravitational lensing of Bozza \cite{bozza2}. \medskip{}

We consider the geometry as follows: A light ray from a source $\left(S\right)$
is deflected by the wormhole acting as a lens $\left(L\right)$ and
reaches the observer $\left(O\right)$. The background spacetime is
asimptotically flat. The line joining lens and observer $\left(OL\right)$
is the optic axis. $\beta$ and $\theta$ are the angular position
of the source and the image with respect to the optical axis, respectively.
The distances between observer and lens, lens and source, and observer
and source are $D_{OL}$, $D_{LS}$ and $D_{OS}$, respectively.

The relation between the position of the source and the position of
the image is called the \emph{lens equation}\cite{ellis},

\begin{equation}
\textrm{tan}\theta-\textrm{tan}\beta=\frac{D_{LS}}{D_{OS}}\left[\textrm{tan}\theta+\textrm{tan}\left(\alpha-\theta\right)\right],\label{lenseq}\end{equation}

where $\alpha$ is the deflection angle. For a spherically symmetric
spacetime with line element

\begin{equation}
ds^{2}=-A\left(x\right)dt^{2}+B\left(x\right)dx^{2}+C\left(x\right)\left(d\theta^{2}+\textrm{sin}^{2}\theta d\phi^{2}\right),\end{equation}

the deflection angle was obtained by Virbhadra et. al. \cite{Vir98} as a function of closest approach $x_{o}=\frac{r_{o}}{2M}$ as

\begin{equation}
\alpha\left(x_{o}\right)=I\left(x_{o}\right)-\pi,\end{equation}

where

\begin{equation}
I\left(x_{o}\right)=\int_{x_{o}}^{\infty}\frac{2\sqrt{B\left(x\right)}dx}{\sqrt{C\left(x\right)}\sqrt{\frac{C\left(x\right)A\left(x_{o}\right)}{C\left(x_{o}\right)A\left(x\right)}-1}}.\end{equation}

The method of Bozza is to expand the integral in $I\left(x_{o}\right)$
around the photon sphere to obtain a logarithmic expression for $\alpha\left(x_{o}\right)$,
and then, by using the lens equation, obtain the position of the produced
images.

\section{Deflection Angle for Wormholes}

The way to calculate the deflection angle produced by a wormhole depends
on the closest approach of the light ray:

\subsection{Closest Approach outside Wormhole's Mouth}

When the closest approach of the light ray is greater than wormhole's
mouth ($r_{o}\geq a$), the deflection angle is produced only by the
external metric. Therefore, the lensing effect is, in this case, exactly
the same as the produced by a Schwarzschild metric with the correspondient
mass. This angle is given \cite{bozza2} by

\begin{equation}
\alpha\left(x_{o}\right)=-2\textrm{ln}\left(\frac{2x_{o}}{3}-1\right)-0.8056.\end{equation}

\subsection{Closest Approach inside Wormhole's Mouth}

Since wormhole solutions described above are defined in two regions,
when the closest approach of the light ray is inside wormhole's mouth
($r_{m}<r_{o}<a$), the deflection angle must be calculated as two
contributions: first a deflection produced by the external Schwarzschild-like
metric, and second the contribution produced by the internal metric.

In both of the specific solutions given, the external contribution
to the deflection angle is given as Schwarzschild deflection above
but with a closest approach equal to the wormhole's mouth,

\begin{equation}
\alpha_{e}=-2\textrm{ln}\left(\frac{2a}{3}-1\right)-0.8056.\label{extern}\end{equation}

The contribution of the internal metric has to be calculated for each
solution:

\subsubsection{Solution A}

For the specific solution A described above we have

\begin{equation}
I_{A}\left(x_{o}\right)=2\int_{x_{o}}^{a}\frac{1}{\sqrt{x^{2}\left(1-\sqrt{\frac{x_{m}}{x}}\right)}}\frac{dx}{\sqrt{\frac{x^{2}}{x_{o}^{2}}-1}}.\end{equation}

Making the change $y=\frac{x}{x_{o}}$ we obtain

\begin{equation}
I_{A}\left(x_{o}\right)=2\int_{1}^{\frac{a}{x_{o}}}\frac{1}{\sqrt{y^{2}\left(1-\sqrt{\frac{x_{m}}{x_{o}y}}\right)}}\frac{dy}{\sqrt{y^{2}-1}}.\end{equation}

The integrand diverges for $y=0$, $y=\frac{x_{m}}{x_{o}}$ and $y=1$.
However, only the third of this values is in the integration range.
Because of this, we will expand the argument of the square root around
this divergence up to the second order in $y$. This gives

\begin{equation}
I_{A}\left(x_{o}\right)=2\int_{0}^{\frac{a}{x_{o}}-1}\frac{dz}{\sqrt{\zeta_{A}z+\eta_{A}z^{2}}},\label{int1}\end{equation}

where

\begin{eqnarray}
z & = & y-1\\
\zeta_{A} & = & 2\left(1-\sqrt{\frac{x_{m}}{x_{o}}}\right)\\
\eta_{A} & = & 2\left(5-4\sqrt{\frac{x_{m}}{x_{o}}}\right).\end{eqnarray}

When $\zeta_{A}$ is non-zero, the leading order of the divergence
is $z^{1/2}$, which can be integrated to give a finite result. When
$\zeta_{A}$ vanishes, the divergence is $z^{-1}$ which makes the
integral to diverge. In this way, the condition $\zeta_{A}=0$ give
us the radius of the photon sphere,

\begin{equation}
x_{ps}=x_{m}.\end{equation}

This means that any photon with a closest approach distance $x_{o}=x_{m}$
will be capturated and it is interesant to note that the radius of
the photon sphere does correspond to the throat radius. Therefore,
any photon with a closest approach distance less than the throat radius
does not emerge in the same part of the universe, but goes into the
wormhole to the other mouth.

Integral in (\ref{int1}) can be exactly made as

\begin{equation}
I_{A}\left(x_{o}\right)=-\frac{2}{\sqrt{\eta_{A}}}\textrm{ln}\left[\frac{\zeta_{A}}{\zeta_{A}+2\eta_{A}\left(\frac{a}{x_{o}}-1\right)+2\sqrt{\eta_{A}\left(\frac{a}{x_{o}}-1\right)\left(\zeta_{A}+\eta_{A}\left(\frac{a}{x_{o}}-1\right)\right)}}\right].\end{equation}

The total deflection angle for solution A is then

\begin{equation}
\alpha_{A}\left(x_{o}\right)=\alpha_{e}+I_{A}\left(x_{o}\right).\end{equation}

\subsubsection{Solution B}

For the specific solution B, we have

\begin{equation}
I_{B}\left(x_{o}\right)=2\int_{x_{o}}^{a}\frac{1}{\sqrt{x^{2}\left(1-\frac{x_{m}^{2}}{x^{2}}\right)}}\frac{dx}{\sqrt{\frac{x^{2}}{x_{o}^{2}}-1}}.\end{equation}

This integral can be rewritten as

\begin{equation}
I_{B}\left(x_{o}\right)=2\int_{1}^{\frac{a}{x_{o}}}\frac{dy}{\sqrt{y^{2}\left(1-\frac{x_{m}^{2}}{x_{o}^{2}y^{2}}\right)\left(y^{2}-1\right)}},\end{equation}

where $y=\frac{x}{x_{o}}$. Once more, the integrand diverges for
$y=0$, $y=\frac{x_{m}}{x_{o}}$ and $y=1$, but we will consider
only the third of this values. Expanding the argument of the square
root around this divergence up to the second order in $y$ we have

\begin{equation}
I_{B}\left(x_{o}\right)=2\int_{0}^{\frac{a}{x_{o}}-1}\frac{dz}{\sqrt{\zeta_{B}z+\eta_{B}z^{2}}},\label{int2}\end{equation}

where this time

\begin{eqnarray}
z & = & y-1\\
\zeta_{B} & = & 2\left(1-\frac{x_{m}^{2}}{x_{o}^{2}}\right)\\
\eta_{B} & = & 2\left(5-\frac{x_{m}^{2}}{x_{o}^{2}}\right).\end{eqnarray}

The condition $\zeta_{B}=0$ give us the radius of the photon sphere,
and again we have

\begin{equation}
x_{ps}=x_{m}\end{equation}

Therefore the behavior of a photon near this radius is similar to
the described before. The integral in (\ref{int2}) can be exactly
made as

\begin{equation}
I_{B}\left(x_{o}\right)=-\frac{2}{\sqrt{\eta_{B}}}\textrm{ln}\left[\frac{\zeta_{B}}{\zeta_{B}+2\eta_{B}\left(\frac{a}{x_{o}}-1\right)+2\sqrt{\eta_{B}\left(\frac{a}{x_{o}}-1\right)\left(\zeta_{B}+\eta_{B}\left(\frac{a}{x_{o}}-1\right)\right)}}\right],\end{equation}

and the total deflection angle for solution B is

\begin{equation}
\alpha_{B}\left(x_{o}\right)=\alpha_{e}+I_{B}\left(x_{o}\right).\end{equation}

\section{Comparison of deflection angles}

As we have shown in the preceeding section, wormholes present an interesting
lensing effect. Photons with a closest approach distance greater than
wormhole's mouth have a Schwarzschild lensing effect, while photons
with a closest approach distance less than wormhole's mouth have a
Schwarzschild lensing plus an inner lensing effect. Because of this,
is important to compare the deflection angle of wormholes with the
deflection of Schwarzschild,

\begin{center}
\includegraphics[scale=0.8]{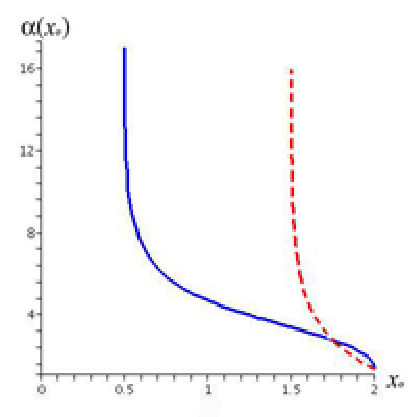}\includegraphics[scale=0.8]{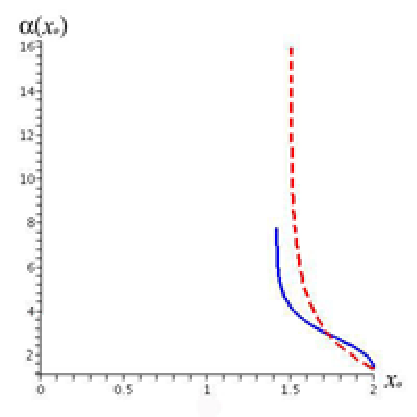}
\par\end{center}

\begin{center}
{\footnotesize Figure 1. In the left side we have the deflection angle
for the wormhole solution A (continuous) compared with Schwarzschild's
deflection angle (dotted). Wormhole's mouth is at $a=2$ (in Schwarzschild
units), and the throat is at $x_{m}=\frac{1}{2}$. It is easy to see
that the blue curve diverges at $x_{o}=x_{m}$. In the right side
we hae the deflection angle for the wormhole solution B (continuous)
compared with Schwarzschild's deflection angle (dotted). Wormhole's
mouth is again at $a=2$ (in Schwarzschild units) and the throat is
at $x_{m}=\sqrt{2}$. The blue curve diverges exactly at $x_{o}=x_{m}$.
In both cases, outside wormhole's mouth the angles coincide.}
\par\end{center}{\footnotesize \par}

\medskip{}

From these graphics we can see that the deflection angle for light
passing outside the wormhole is the same as the one produced by Schwarzschild,
and some differences appear when light goes inside. One of the most
important is the divergence in the blue curves: they occur exactly
at wormhole's throat, showing that this surface corresponds to a photon
sphere.

\section{Magnification}

Magnification is an important characteristic of Gravitational Lensing
because it is easily observable. As is already known, magnification
is given by

\begin{equation}
\mu=\frac{\mbox{sin}\beta}{\mbox{sin}\theta}\frac{d\beta}{d\theta}.\end{equation}

Then, in order to obtain the magnification produced by a wormhole
we will use the produced deflection angle and, using the lens equation
(\ref{lenseq}), obtain a relation between the angular positions of
the image $\left(\theta\right)$ and the source $\left(\beta\right)$.
since the deflection angle is calculated for each of the presented
solutions, the magnification must be evaluated also for each wormhole.

\subsection{Solution A}

For the first of the presented solutions the magnification is given
by:

\begin{equation}
\mu=\left(1-\frac{D_{ds}D_{d}}{D_{s}}\sqrt{1-\sqrt{\frac{x_{m}}{a}}}\frac{\alpha\left(x_{o}\right)}{x_{o}}\right)^{-1}\left(1-\frac{D_{ds}D_{d}}{D_{s}}\sqrt{1-\sqrt{\frac{x_{m}}{a}}}\frac{d\alpha}{dx_{o}}\right)^{-1},\end{equation}

where $\alpha\left(x_{o}\right)$ is the deflection obtained before
and the derivative $\frac{d\alpha}{dx_{o}}$ can be evaluated using
a math software. If we plot the magnification as a function of the
closest aproach we have the following curves.

\begin{center}
\includegraphics[scale=0.5]{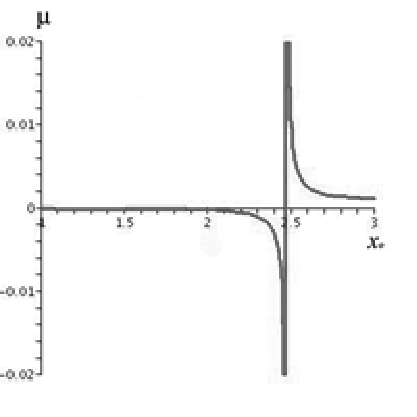}\includegraphics[scale=0.5]{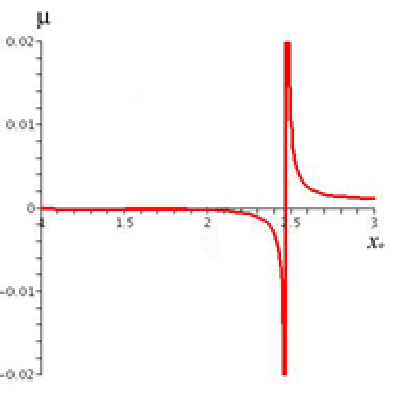}
\par\end{center}

\begin{center}
{\footnotesize Figure 2. Comparison of the magnification for Schwarzschild's
black hole (left) and wormhole A (right). At this scale the curves
look the same. Wormhole's mouth is at $a=2$}
\par\end{center}{\footnotesize \par}

\medskip{}

As it is easily seen, the two curves look exactly the same. Moreover,
we can see the divergence that corresponds to the Einstein' ring as
is expected because the mouth of the wormhole is situated inside this
ring and the exterior metric is exactly the same as Schwarzschild's
metric.

However if we made a zoom near the wormhole's mouth, we can see how
in the wormhole's case, the magnification shows a discontinuity:

\begin{center}
\includegraphics[scale=0.4]{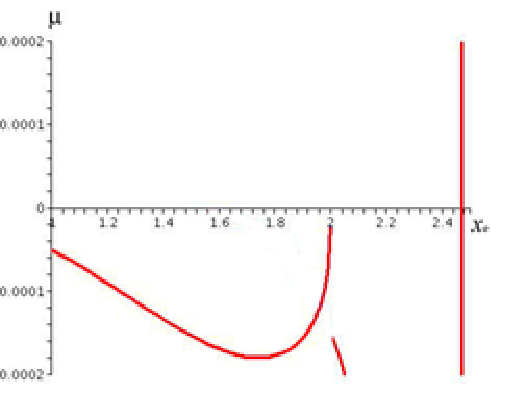}
\par\end{center}

\begin{center}
{\footnotesize Figure 3. Zoom of the magnification for wormhole A,
near its mouth located at $a=2$}
\par\end{center}{\footnotesize \par}

\medskip{}

It is important to note that the discontinuity occurs exactly at the
mouth of the wormhole, and it happens because here is were the matter
distribution of the wormhole begins. It is also seen a little peak
in the magnification curve inside the wormhole that is not seen in
Schwarzschild geometry.

Moreover, when we change the wormhole's parameters, a very interesting
behavior appears. In the next set of figures we can see the magnification
curve for wormholes with different mouth sizes. 

\begin{center}
\includegraphics[scale=0.35]{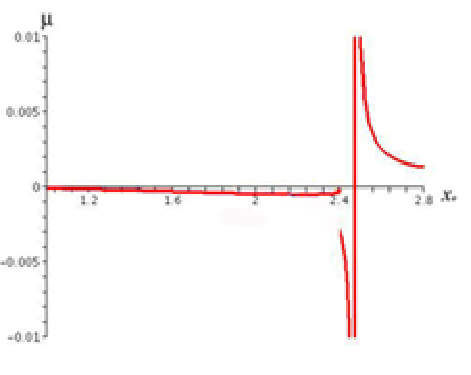}\includegraphics[scale=0.35]{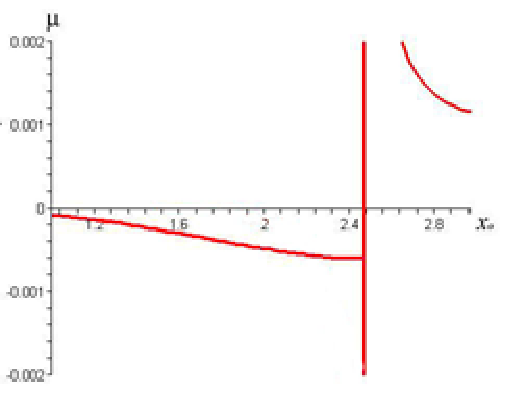}
\par\end{center}

\begin{center}
\includegraphics[scale=0.35]{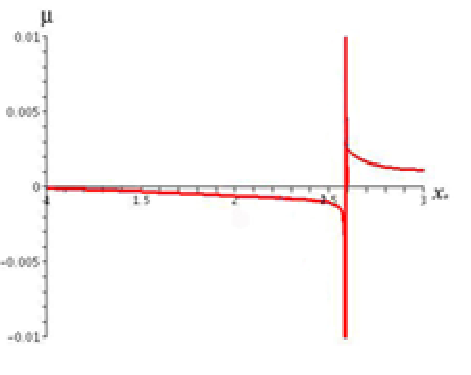}\includegraphics[scale=0.35]{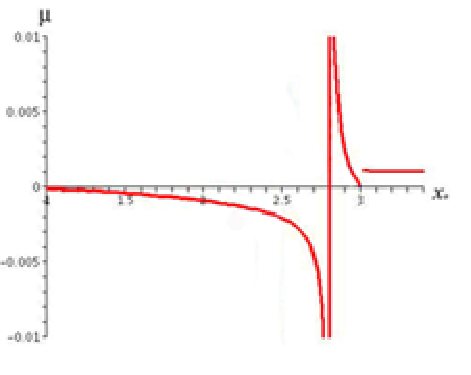}
\par\end{center}

\begin{center}
{\footnotesize Figure 4.Magnification curves for wormhole A with different
mouth's sizes. In the first we have $a=2.4$, in the second $a=2.47$,
in the third $a=2.6$ and in the fourth $a=3$.Note that in each case
we ohave a discontinuity at $a$.}
\par\end{center}{\footnotesize \par}

\medskip{}

It is easily seen, that when we change the mouth of the wormhole,
the little peak inside grows. When the mouth of the wormhole is inside
the Einstein's ring radius the magnification behaves just as commented
before, but when the mouth of the wormhole is greater that the Einstein
ring the peak in magnification of this ring dissapears and there appears
a new peak produced by the wormhole. Observationally, this fact shows
an {}``Einstein's ring'', shifted from the original position and
this effect can be, in principle, measured.

\subsection{Solution B}

For wormhole B we obtain a magnification given by the expression

\begin{equation}
\mu=\left(1-\frac{D_{ds}D_{d}}{D_{s}}\sqrt{1-\frac{x_{m}^{2}}{a^{2}}}\frac{\alpha\left(x_{o}\right)}{x_{o}}\right)^{-1}\left(1-\frac{D_{ds}D_{d}}{D_{s}}\sqrt{1-\frac{x_{m}^{2}}{a^{2}}}\frac{d\alpha}{dx_{o}}\right)^{-1}.\end{equation}

Again, we have here the deflection angle for solution B, $\alpha\left(x_{o}\right)$
and the derivative $\frac{d\alpha}{dx_{o}}$, that is evaluated using
a math software. If we plot the magnification as a function of the
closest approach $x_{o}$, the curve looks exactly the same as the
Schwarzschild magnification (just as for the wormhole soultion a in
Figure 2.). However, when we make a zoom of the region near wormhole's
mouth we get a little difference:

\begin{center}
\includegraphics[scale=0.4]{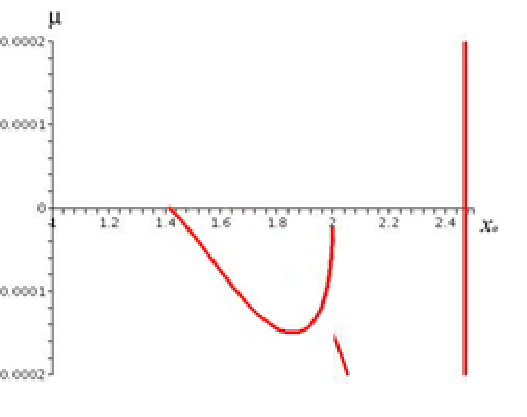}
\par\end{center}

\begin{center}
{\footnotesize Figure 5. Zoom of the magnification for wormhole B,
near its mouth located at $a=2$}
\par\end{center}{\footnotesize \par}

\medskip{}

See how we have again a discontinuity localizated at the wormhole's
mouth and a little peak inside the wormhole that makes the diference
when compared with Schwarzchild's geometry.Just as in the specific
solution B, the inside peak's position depends on the size of the
wormhole's mouth. This fact is shown in the next set of figures:

\begin{center}
\includegraphics[scale=0.35]{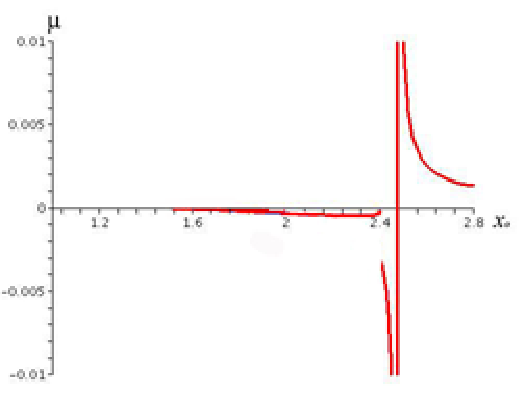}\includegraphics[scale=0.35]{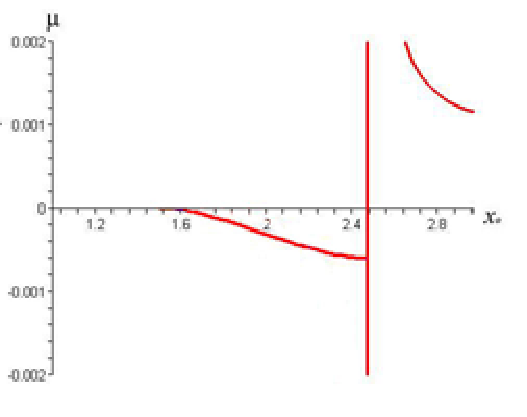}
\par\end{center}

\begin{center}
\includegraphics[scale=0.35]{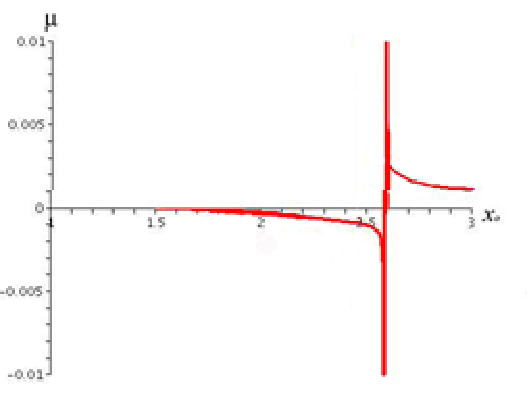}\includegraphics[scale=0.35]{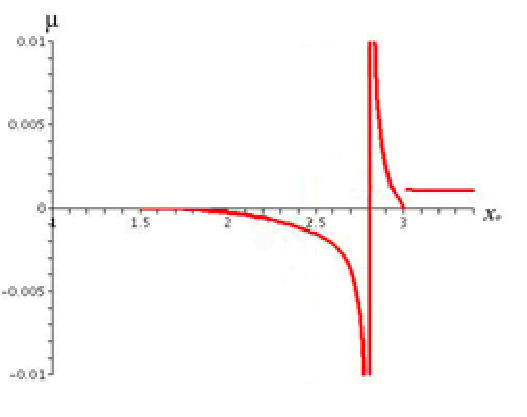}
\par\end{center}

\begin{center}
{\footnotesize Figure 6.Magnification curves for wormhole B with different
mouth's sizes. In the first we have $a=2.4$, in the second $a=2.47$,
in the third $a=2.6$ and in the fourth $a=3$.Note that in each case
we ohave a discontinuity at $a$.}
\par\end{center}{\footnotesize \par}

\medskip{}

Again it is seen an observational difference in the magnification
because there is a peak that does not correspond exactly to the Einstein's
ring but to the wormhole with a sufficiently greater mouth.

\section{Conclusion}

Gravitational lensing is one of the most important effects in General
Relativity. In this paper we have applied the strong field lensing
method to Morris-Thorne wormholes and the analysis of the deflection
angle and magnification produced by the two specific wormhole solutions
described show that the behaviour of these objects is very similar
to the behaviour of Schwarzschild's solution. As can be seen from
the graphics, there is a logarithmic divergence of the deflection
angle, and there is a photon sphere that corresponds to the throat
of the wormhole. This behavior contrast with the behaviour reported
by Cramer et. al. \cite{cramer} and Safanova et. al. \cite{safanova}
since their model for the wormhole is a negative punctual mass. Altough
the two solutions considered here need some exotic (gravitationally
negative) mass inside, the behavior of light deflection does not differ
significatively from the Schwarzschild case. 

On the other side, magnification curves for the wormholes are similar
to the obtained for Schwazschild, but when looking in some detail,
there is a discontinuity in the curve (at the point where the mass
distribution begins). It is also of great importance to note here
that the size of the mouth of the wormhole affects the behavior of
the magnification curve, and specially when the mouth of the wormhole
is greater than the size of the Einstein's ring. In this last case
we see how a ring appears but it is not at the expected position for
the Einstein's ring, making a possible way to identify wormholes from
other mass distributions as, for example, black holes.

\end{document}